\documentclass[a4paper,11pt]{article}
\usepackage{pos}

\usepackage[sort&compress]{natbib}
\setcitestyle{numbers}
\usepackage{caption}
\usepackage{wrapfig}
\usepackage{physics}
\usepackage{amsmath}
\usepackage{array,mathtools,booktabs}
\usepackage{bbold}
\usepackage{siunitx}
\usepackage{tabu}
\definecolor{linkblue}{HTML}{0066cc}
\usepackage{tikz}
\usetikzlibrary{positioning}
\usepackage{cleveref}

\graphicspath{
	{.} 
	{./figs/}
}

\title{
Progress in lattice simulations for two Higgs doublet models }

\renewcommand{\printHeadAuthors}{G. Catumba}
\author*[a]{Guilherme Catumba}
\author[b,g]{Atsuki Hiraguchi}
\author[c]{George W.-S Hou}
\author[d]{Karl Jansen}
\author[c,e]{Ying-Jer Kao}
\author[b,f]{C.-J. David Lin}
\author[a]{Alberto Ramos}
\author[b,c]{Mugdha Sarkar}

\affiliation[a]{Instituto de Física Corpuscular (IFIC) CSIC - Universitat de Valencia. \\ 46071, Valencia, Spain}
\affiliation[b]{Institute of Physics, National Yang Ming Chiao Tung University, 1001 Ta-Hsueh Road, Hsinchu 30010, Taiwan}
\affiliation[c]{Department of Physics, National Taiwan University, Taipei 10617, Taiwan}
\affiliation[d]{Deutsches Elektronen-Synchrotron DESY, Platanenallee 6, 15738 Zeuthen, Germany} \affiliation[e]{Center for Theoretical Physics and Center for Quantum science and technology, National Taiwan University, Taipei, 10607, Taiwan}
\affiliation[f]{Center for High Energy Physics, Chung-Yuan Christian University, Chung-Li 32023, Taiwan}
\affiliation[g]{CCSE, Japan Atomic Energy Agency, 178-4-4, Wakashiba, Kashiwa, Chiba 277-0871, Japan}

\emailAdd{gtelo@ific.uv.es}

\abstract{

The custodial Two-Higgs-Doublet-Model with SU(2) gauge fields is studied on the
lattice. This model has the same global symmetry structure as the Standard Model
but the additional Higgs field enlarges the scalar spectrum and opens the
possibility for the occurrence of spontaneous symmetry breaking of the global
symmetries. Both the spectrum and the running of the gauge coupling of the
custodial 2HDM are studied on a line of constant Standard Model physics with
cutoff ranging from 300 to 600 GeV. The lower bounds of the realizable masses
for the additional BSM scalar states are found to be well bellow the W boson
mass. In fact, for the choice of quartic couplings in this work the estimated
lower mass for one of the BSM states is found to be about $\sim 0.2m_{W}$ and
independent of the cutoff.

}

\FullConference{%
The 41st International Symposium on Lattice Field Theory,\\
 July 28th to August 3rd, 2024,\\
University of Liverpool, United Kingdom
}

\graphicspath{
  {.} 
}

\begin{document}
\maketitle

\section{Introduction} \label{sec:Introduction}

The standard model (SM) has been very successful in explaining experimental
results.  However, searches for physics beyond the SM (BSM) are important and
should be pursued, with one of the key motivations being the need of a strong
first-order electroweak phase transition (EWPT) in the theory of electroweak
baryogenesis.  It has been known that the SM does not facilitate such a phase
transition at the measured Higgs boson mass, $m_{h} \approx 125$ GeV
\cite{Evertz:1986af,D_Onofrio_2016}.

The simplest BSM extension complying with the observations is the
\textit{two-Higgs-doublet model} (2HDM), where a second $SU(2)$-doublet scalar
field is added to the theory.  This model yields an enlarged spectrum and also
rich phenomenology.  Dark matter candidates within the 2HDM-type, usually called
\textit{inert double models}, were proposed in
\cite{Ma:2006km,Barbieri_2006,honorez_inert_2007}.  Moreover, 2HDMs are also an
important part of many supersymmetric theories such as the Minimal
Supersymmetric Standard Model
\cite{Haber:1984rc,haber_renormalization-group-improved_1993}.

The Lagrangian of the 2HDM considered in this work contains the kinetic terms of
both scalar and gauge fields, and the most general $SU(2)$-invariant scalar
potential. Since we are interested in the lattice application, from now on we
consider the case of real couplings only.  This is the CP-conserving 2HDM
\cite{Haber_2011,ONeil:2009fty}.  With this choice, the most general
renormalizable 2HDM scalar potential has 10 real parameters, and in the usual
doublet formulation reads
\begin{equation}
  \label{eq:cont_potential}
\begin{aligned}
  V_{\text{2HDM}} &= \mu_{11}^{2}\phi_{1}^{\dagger}\phi_{1} + \mu_{22}^{2}\phi_{2}^{\dagger}\phi_{2} + \mu_{12}^{2}\Re (\phi_{1}^{\dagger}\phi_{2}) + \lambda_1 (\phi_{1}^{\dagger}\phi_{1})^2 + \lambda_2 (\phi_{2}^{\dagger}\phi_{2})+ \lambda_3(\phi_{1}^{\dagger}\phi_{1})(\phi_{2}^{\dagger}\phi_{2})\\
  &+ \lambda_4(\phi_{1}^{\dagger}\phi_{2})(\phi_{2}^{\dagger}\phi_{1}) + \lambda_{5} \Re(\phi_{1}^{\dagger}\phi_{2})^{2} +\Re(\phi_{1}^{\dagger}\phi_{2})\left[ \lambda_{6} (\phi_{1}^{\dagger}\phi_{1}) + \lambda_{7} (\phi_{2}^{\dagger}\phi_{2})\right].
\end{aligned}
\end{equation}

The amount of literature on the 2HDM is vast and on-going (see ref.
\cite{branco_theory_2012} and references therein).  Most of the works focus on a
perturbative, tree-level analysis of the theory.  On the other hand, the lattice
studies of the 2HDM are scarce \cite{lewis_spontaneous_2010,maas2014observables}
and a thorough non-perturbative analysis is needed.  Furthermore, recent
investigations suggest that large self interacting couplings in the BSM sector
may be realizable within the SM bounds \cite{HOU1992179,Hou_2018}. In fact,
finite-temperature studies indicate that baryogenesis in 2HDM may benefit from
large couplings
\cite{Fromme:2006cm,dorsch_strong_2013,basler_strong_2017,bernon_new_2018}, such
that the electroweak phase transition is of strong first-order in this regime.
Two aspects are of importance: the study of the spectrum of the theory, and the
non-perturbative analysis of the finite-temperature transition, both of which
have not been explored by lattice simulations, and are covered in this work.

In this work we take the $\mathbb Z_{2}$-breaking terms in
\cref{eq:cont_potential} to vanish, $\mu_{12}=\lambda_{6}= \lambda_{7} = 0$ ,
defining the so-called \textit{inert models}. Moreover, by imposing the
condition $\lambda_{4}=\lambda_{5}$ we obtain the custodial limit of the 2HDM.
In this case the symmetry structure is identical to the SM, with the action being
invariant under global $SU(2)$ transformations where both fields transform
simultaneously.

The lattice action for the custodial 2HDM used in this work can then be written as
\begin{align}
  \label{eq:lattice action}
  S_{\text{2HDM}}&= \sum_{x}\sum_{n=1}^{2} \bigg\{ \sum_{\mu} -2 \kappa_{n}\Tr \left( \Phi_{n}^{\dagger}U_{\mu}\Phi_{n}(x+\mu) \right)+  \Tr \left( \Phi_{n}^{\dagger}\Phi_{n} \right) + \eta_{n}\left[ \Tr \left( \Phi_{n}^{\dagger}\Phi_{n} \right) - 1 \right]^{2}\bigg\}  \nonumber \\
  &  + 2\mu^{2}\Tr \left( \Phi_{1}^{\dagger}\Phi_{2} \right) + \eta_{3}\Tr \left( \Phi_{1}^{\dagger}\Phi_{1} \right)\Tr \left( \Phi_{2}^{\dagger}\Phi_{2} \right) + 2\eta_{4}\Tr \left( \Phi_{1}^{\dagger}\Phi_{2} \right)^{2} + S_{\text{YM}},
\end{align}
where
$S_{\text{YM}} = \beta\sum_{x}\sum_{\mu>\nu}\left[ 1 - \frac{1}{2}\Re \Tr U_{\mu\nu}(x) \right]$
is the standard Wilson plaquette action with $U_{\mu\nu}$ being the plaquette
and $\beta = 4/g^{2}$.

In this formulation we consider the quaternion formalism, where scalar fields in
the fundamental representation of $SU(2)$ are written as a matrix,
$\Phi_{n}(x) = \frac{1}{\sqrt{2}}\sum_{a=1}^{4}\theta_{\alpha}\varphi_{n}^{\alpha}(x)$
with $\varphi^{\alpha}_{n}$ being real components, $\alpha=1,2,3,4$,
$\theta^{4}=\mathbb{1}_{2\times2}$, and $~\theta^{k}=i\sigma^{k}$ for $k=1,2,3$
with $\sigma^{k}$ being the Pauli matrices.  In this representation the fields
$\Phi_{n}$ transform under the $SU(2)$ gauge group by a left multiplication, and
under the global $SU(2)$ by right multiplication,
$\Phi_{n}\rightarrow L(x)\Phi_{n} R$.

\section{Phase structure \& spectrum} \label{sec:phase_structure}

While the single Higgs doublet model, governed by three bare couplings, has a
simple phase structure\footnote{In fact, only a single phase exists in the
Higgs-gauge interaction. The confinement and Higgs phases are analytically
connected.}, the enlarged parameter space of the 2HDM complicates the phase
structure.

\begin{figure}[htb!]
  \centering
\begin{tikzpicture}[very thick]
    \definecolor{salmon}{RGB}{250, 128, 114}
    \def\size{3.4cm}
    \node[fill=lightgray!30,draw=black!30, minimum size=\size, inner sep=0pt] (H0) {};
    \node[fill=white,draw=black!30,minimum width=\size, minimum height=\size, inner sep=0pt, above=-\pgflinewidth of H0] (H2) {};
    \node[fill=white,draw=black!30,minimum width=\size, minimum height=\size, inner sep=0pt, right=-\pgflinewidth of H0] (H1) {};
    \node[fill=lightgray!30,draw=black!30, minimum size=\size, inner sep=0pt,  above right=-\pgflinewidth and -\pgflinewidth of H0] (H12) {};

    \node[anchor=east] at (H0.north west) {$\kappa_{2}^{\rm c}$};
    \node[anchor=east] at (H2.north west) {$\kappa_{2}$};
    \node[anchor=south west] at (H0.south west) {\textcolor{salmon}{$(H_{0})$}};
    \node[anchor=north west] at (H2.north west) {\textcolor{salmon}{$(H_{2})$}};
    \node[anchor=north] at (H0.south east) {$\kappa_{1}^{\rm c}$};
    \node[anchor=north] at (H1.south east) {$\kappa_{1}$};
    \node[anchor=south east] at (H1.south east) {\textcolor{salmon}{$(H_{1})$}};
    \node[anchor=north east] at (H12.north east) {\textcolor{salmon}{$(H_{12})$}};

    \node[yshift=-.5cm,anchor=north] at (H2.north) {$SU(2)\times (\mathbb Z_{2})^{2}$};
    \node[yshift=-.5cm,anchor=north] at (H1.north) {$SU(2)\times (\mathbb Z_{2})^{2}$};
    \node[yshift=-.5cm,anchor=north] at (H0.north) {$SU(2)\times (\mathbb Z_{2})^{2}$};

    \footnotesize
    \node[yshift=-.5cm,xshift=-.3cm,anchor=center] at (H0.center)
    {\begin{varwidth}{\linewidth}\begin{itemize}
        \setlength\itemsep{-0em}
        \item[-] QCD-like
        \item[-] $m_{1^{-}}>m_{0^{+}}$
    \end{itemize}\end{varwidth}};

    \node[yshift=-.5cm,xshift=-.3cm,anchor=center] at (H2.center)
    {\begin{varwidth}{\linewidth}\begin{itemize}
        \setlength\itemsep{-0em}
        \item[-] Degenerate $W$-boson
        \item[-] SM Higgs $m_{h}$
        \item[-] BSM scalar $m_{H}$
        \item[-] 3 degenerate BSM\\ scalars $m_{A}= m_{H^{\pm}}$
    \end{itemize}\end{varwidth}};

    \node[yshift=-.5cm,xshift=-.0cm,anchor=center] at (H1.center)
    {\centering (similar to $H_{2}$)};

    \node[yshift=-.5cm,xshift=-.3cm,anchor=center] at (H12.center)
    {\begin{varwidth}{\linewidth}\begin{itemize}
        \setlength\itemsep{-0em}
        \item[-] 3 non-degenerate $1^{-}$
        \item[-] 3 Goldstone Bosons
        \item[-] 2 scalar states\\ $m_{h}$, $m_{H}$
    \end{itemize}\end{varwidth}};


\end{tikzpicture}
\caption{Summary of the cutodial 2HDM parameter space with the global symmetries
and spectrum content for each of the sectors $H_{0}$, $H_{1}$, $H_{2}$,
$H_{12}$.}
\label{fig:scheme_sectors}
\end{figure}

In the case of the inert models, it is possible to divide the parameter space in
four different regions where none, one, or both scalar fields are in the Higgs
phase \cite{branco_theory_2012}. In the perturbative formulation, using the
vacuum expectation values, $v_{i}$, for each scalar field this corresponds to:
$(H_{0}): v_{1}=v_{2}=0$, $(H_{2}): v_{1}=0, v_{2}\neq0$,
$(H_{1}): v_{1}\neq0, v_{2}=0$, $(H_{12}): v_{1}\neq0, v_{2}\neq0$.  These
sectors divide the $\kappa_{1}, \kappa_{2}$ plane into four regions defined by
the critical values, $\kappa_{i}^{c}$.  Note, however, that not all four can be
seen as separate phases since some of these transitions are crossovers for
certain regimes of the couplings.  The global symmetries and predicted particle
content of each of the phases are summarized in \cref{fig:scheme_sectors}.

Sectors $(H_{1}),(H_{2})$, and $(H_{12})$ have the Higgs mechanism active.
However, the custodial $SU(2)$ symmetry is spontaneously broken in $(H_{12})$.
The absence of the custodial symmetry, and the presence of massless Goldstone
bosons in the spectrum exclude this phase from phenomenological considerations,
since a viable 2HDM has to reproduce SM physics at low energies.

Sectors $(H_{1})$ and $(H_{2})$ are equivalent, and for definitiveness we will
work with the latter.  This is defined by $\kappa_{2}>\kappa_{2}^{c}$,
$\kappa_{1}<\kappa_{1}^{c}$, and consequently, it is the field $\Phi_{2}$ that
reproduces the SM-like Higgs.  At tree-level this model predicts the existence
of three degenerate scalar particles,
$m_{A}^{2} = m_{H^{\pm}}^{2}=\mu_{11}^{2}+\lambda_{3}v^{2}/2$, and two other
scalar states, $m_{H}^{2}=m_{H^{\pm}}^{2} + \lambda_{4}v^{2}$, and
$m_{h}^{2}=\mu_{22}^{2}v^{2}$, the latter being identified with the SM Higgs.

On the lattice the spectrum is obtained in sector $(H_{2})$ by studying the
large time behavior of two-point functions
$\expval{\mathcal O(t)\mathcal O(0)}$, where $\mathcal O$ is a zero momentum
composite operator with the quantum number of the particle of interest.  In
particular we investigate the following operators,
\begin{align}
    &S_{ij}^{a}(x^{4}) = \sum_{\vec x} \Tr\left[\Phi_i^\dagger(x)\Phi_j(x)\theta^{\alpha}\right], && W_{ij,\mu}^a(x^{4}) = \sum_{\vec x} \Tr\left[\Phi_i^\dagger(x)U_{\mu}(x)\Phi_j(x+\hat\mu)\theta^{\alpha}\right].\label{eq:interpolators}
\end{align}
In sector $(H_{2})$ the assignment between the interpolators and the spectrum
is: $S_{22}$ sources the SM Higgs, and $W_{22}^{j}$ the W bosons; $S_{12}^{4}$
or $W_{12}^{4}$ source the scalar state $H$, and $S_{12}^{j}$ or $W_{12}^{j}$
the scalar states $A$ and $H^{\pm}$.

\section{Standard model physics} \label{sec:standard_model_physics}

In order to study the cutoff dependence of the enlarged spectrum of the theory
we must define a line of constant physics.  Since we are interested in
performing a non-perturbative study of the BSM scalar states in the 2HDM, and
given that only SM physics is available, we build a line of partially constant
physics (LPCP). The Higgs sector of the SM has two independent dimensionless
renormalized quantities that can be conveniently defined for the LPCP.  Namely,
the ratio of the Higgs to the $W$ boson masses, $R \equiv \frac{m_{h}}{m_{W}}$
and the renormalized running gauge coupling $g_{R}^{2}(\mu)$.    In the
following, we fix the $R$-ratio to be close to the SM value, $R \approx 1.5$,
while the renormalized gauge coupling is set to its physical value at the scale
of the $W$ boson mass, $g_{R}^{2}(\mu=m_{W})\equiv 4\pi a_{W}\sim 0.5$.

The definition of the renormalized running gauge constant is done through the
gradient flow action density \cite{Luscher_2010_flow},
\begin{equation}
  \expval{E(x,t)} = -\frac{1}{4}\expval{G_{\mu\nu}^{a}(x,t)G_{\mu\nu}^{a}(x,t)},
\end{equation}
where
$G_{\mu\nu}(x,t) = \partial_\mu B_\nu(x,t) - \partial_\nu B_\mu(x,t) + \left[ B_\nu(x,t), B_\mu(x,t)  \right]$
is the flowed gauge field strength.  On the lattice we use the Clover
discretization of the field strength tensor and the Wilson action for the
discretized flow equation.

Using the relation between the renormalized gauge coupling at the scale
$\mu = 1/\sqrt{8t}$ and the flowed action density
\cite{luscher_perturbative_2011}, we define the gradient flow renormalized gauge
coupling by
\begin{align}
  \label{eq:GF_coupling}
    g_{GF}^{2}(\mu)\equiv \frac{128\pi^{2}}{9} \eval{t^{2}\expval{E(t)}}_{t=1/8\mu^{2}}, && g_{GF}^{2}(\mu = m_{W}) = 0.5,
\end{align}
where $\expval{E(t)}$ is obtained from the Euclidean spacetime average of
$\expval{E(x,t)}$.  On the lattice this condition is equivalent to
$S\equiv \sqrt{8t_{0}}m_{W} = 1.0$ where the flow scale $t_{0}/a^{2}$ is defined
by \cref{eq:GF_coupling}.

\section{Results} \label{sec:Results}

The adopted strategy to build the LPCP is to sequentially increase the bare
coupling $\beta$ towards the continuum while scanning the
$\{\kappa_{2},\eta_{2}\}$-space to find parameters such that the above SM
conditions are satisfied.

In this strategy the remaining degrees of freedom of the theory are kept fixed,
namely the value of the BSM bare couplings
($\kappa_{1},\eta_{1}, \eta_{3},\eta_{4}$) are chosen such that the additional
masses $m_{H}$ and $m_{A}=m_{H^{\pm}}$ are larger than the SM masses but well
below the lattice cutoff. Finally, the couplings $\eta_{3},\eta_{4}$ were chosen
to be small.  In this way, the LPCP in the SM sector should be mostly
insensitive to the $\Phi_{1}$ scalar sector.  Notice that while the initial
strategy keeps the BSM bare couplings constant, later we will explore the effect
of the BSM couplings.

\begin{table}
  \centering
  \begin{tabular}{cccccc}
    \toprule
 $\beta$                &        8.2 &        8.3 &        8.4 &       8.56 &       8.64 \\
 $\kappa_2$             &    0.13175 &    0.13104 &     0.1306 &     0.1301 &   0.129985 \\
 $\eta_2$               &    0.00338 &      0.003 &    0.00285 &    0.00275 &   0.002737 \\
    \midrule
 $R$                    &  1.509(94) &  1.527(80) &  1.494(65) &  1.504(39) &  1.462(53) \\
 $S$                    & 1.0055(98) & 0.9956(65) &  0.994(14) &  0.994(20) & 0.9958(94) \\
 $am_{h}$               &  0.402(28) &  0.363(21) &  0.305(16) & 0.2192(84) & 0.1863(75) \\
 $am_{W}$               & 0.2666(26) & 0.2377(15) & 0.2041(29) & 0.1458(29) & 0.1275(11) \\
 $t_0/a^2$              & 1.7781(58) & 2.1934(74) &  2.966(13) &  5.810(45) &  7.626(43) \\
 $\Lambda~(\si{GeV})$ &  301.5(29) &  338.2(21) &  393.8(57) &    551(11) &  630.4(56) \\
 $m_{W}L$                   &  8.485(14) &  7.639(13) &  6.569(15) &  4.694(18) &  6.145(17) \\

    \bottomrule
  \end{tabular}
  \caption{Bare couplings $\beta,\kappa_{2},\eta_{2}$ of the LPCP together with
the corresponding physical conditions $R,S$, the SM masses $am_{h},am_{W}$ in
lattice units, the gradient flow scale $t_{0}/a^{2}$ and the estimated cutoff
energy $\Lambda_{c}$. The remaining couplings were fixed to the values:
$\kappa_1=0.1245$; $\eta_1=0.003$; $\eta_{3}= 0.002$; $\eta_{4}=\eta_{5}=0.0001$.
The simulations were performed on lattices with $L=28,28,32,32,48$.}
  \label{tab:LCP}
\end{table}

\begin{figure}[htb!]
  \centering
		\includegraphics[width=0.49\textwidth]{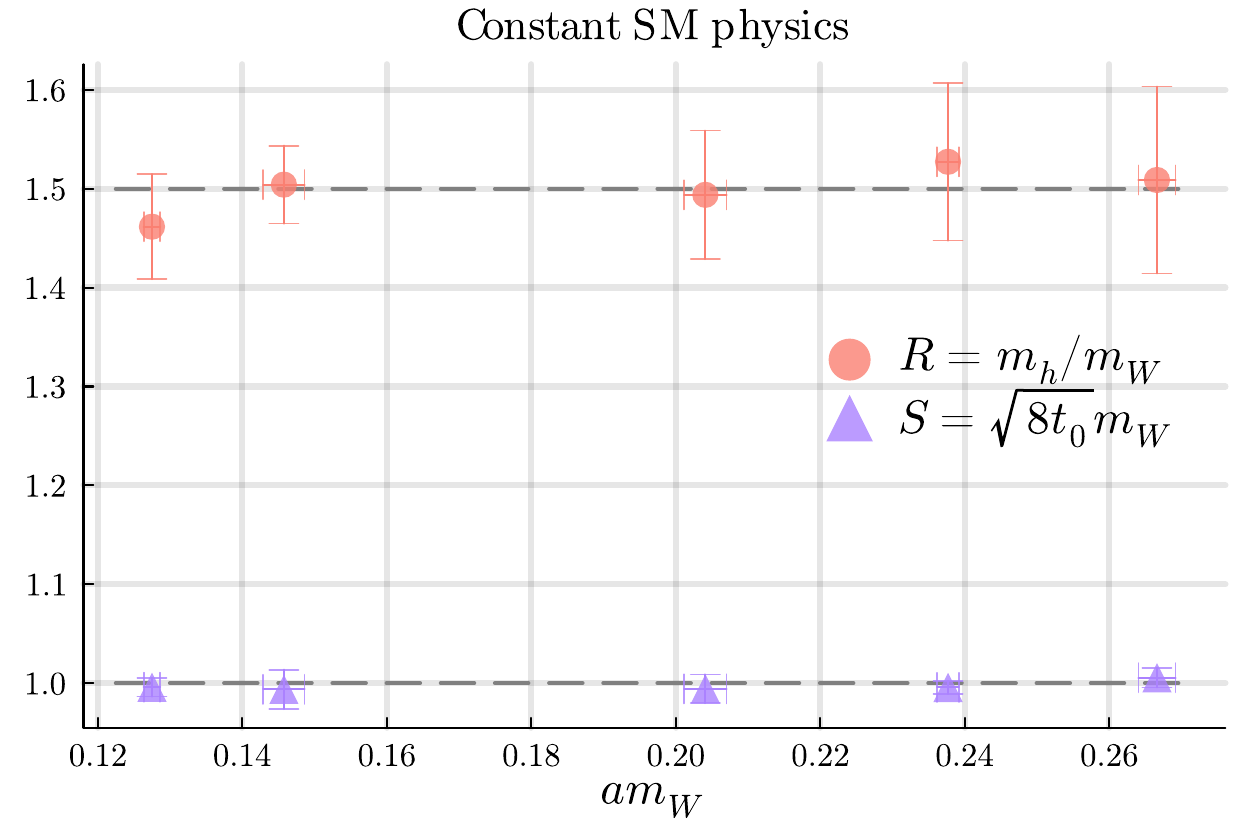}
    \includegraphics[width=0.49\textwidth]{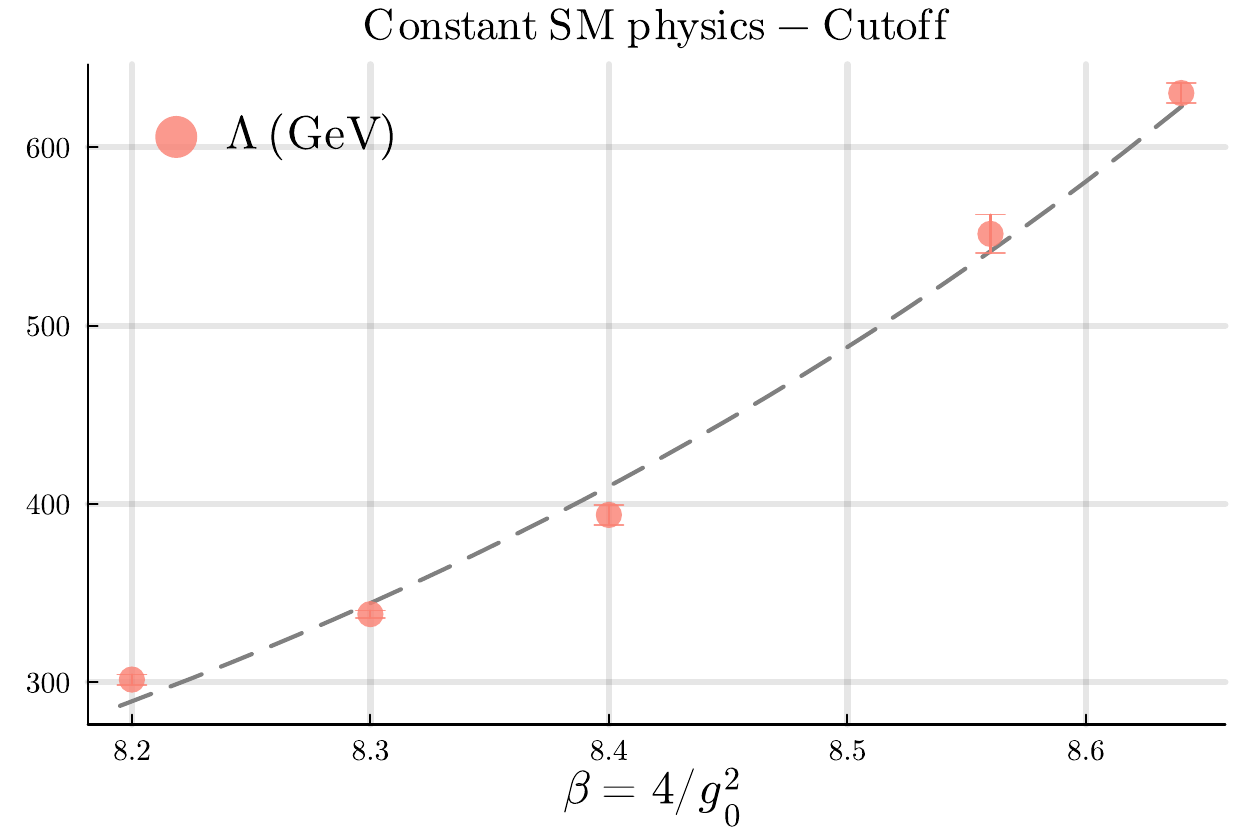}
\caption{Data from \cref{tab:LCP}: physical conditions $R$ and $S$ for the
selected points in the line of constant SM physics as a function of $am_{W}$
(left).  For decreasing $am_{W}$, each points has a corresponding increasing
$\beta$ value.  The lattice cutoff, $\Lambda=1/a$, estimated from
$a=\hat m_{W}/m_{W}^{\textrm{phys}}$ is shown on the right as a function of
$\beta$.  }
\label{fig:lcp}
\end{figure}

The results for the tuning of the LPCP defined by $\beta,\kappa_{2},\eta_{2}$
are shown in \cref{tab:LCP}.  Five $\beta$ values were considered, with the
cutoff ranging from $300~\si{GeV}$ to $630~\si{GeV}$.  The scale setting is
obtained from the physical value $m_{W}^{\textrm{phys}}=80.377(12)~\si{GeV}$
\cite{Workman:2022ynf} with $a=\hat m_{W}/m_{W}^{\textrm{phys}}$, where $\hat m$
is the mass in lattice units.  Both the physical conditions $R,S$, and the
lattice cutoff $\Lambda=1/a$ are shown in \cref{fig:lcp} as a function of
$am_{w}$ and $\beta$, respectively.



\begin{figure}[htb!]
  \centering
  \includegraphics[width=0.7\textwidth]{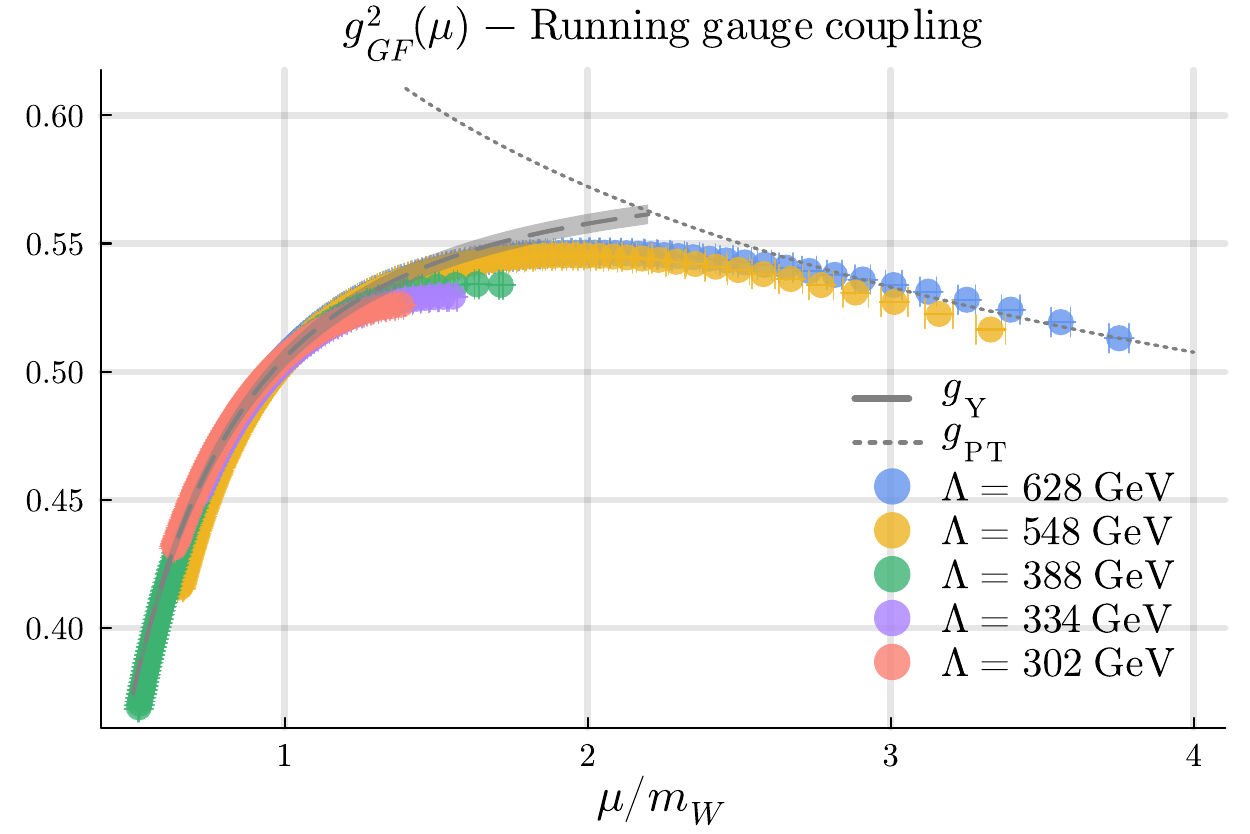}
  \caption{Running gauge coupling $g_{GF}^{2}(\mu)$ as a function of $\mu/m_{W}$
from lattice simulations along the LPCP in \cref{tab:LCP}. Only scales with
$\sqrt{8t}/a>2$ are shown. The perturbative result is matched to the curve
corresponding to $\Lambda=628~\si{GeV}$. The gauge coupling from the Yukawa
potential is fitted to the  infrared region, with the estimated screening mass
$m_{\textrm{screen}}= 0.6243(28)m_{W}$.}
  \label{fig:running_coupling}
\end{figure}

The gradient flow running of the gauge coupling was computed for each point in
\cref{tab:LCP}.  The results for  $g_{GF}^{2}(\mu=1/\sqrt{8t};\beta)$ are shown
in \cref{fig:running_coupling} as a function of the energy scale relative to
$m_{W}$.  All curves match for a large range, with lattice artifacts being only
apparent for large values of $\mu$.\footnote{For each curve, only energies below
a certain threshold are shown, corresponding to flow time radius
$\sqrt{8t}/a>2$, that correspond to $t/a^{2}>0.5$. Smaller flow times lead to
large lattice artifacts.}

In \cref{fig:running_coupling} the perturbative result, $g_{\rm PT}^{2}$ is also
shown for large energies.  This was computed from the one-loop $\beta$-function
$\beta_{SU(N)+\rm Scalars}= \mu \dv{g}{\mu} = -\frac{b_0g^3}{16\pi^2}+\order{g^5}$,
with $b_0=\frac{11N-n_s}{3}$, and $n_{s}=2$ is the number of scalar fields and
$N=2$. The matching between the massive non-perturbative and the massless
perturbative schemes was done at a large enough energy scale,
$\mu \approx 3.5 m_{W}$.

From \cref{fig:running_coupling} the structure of the running can be understood
as follows: for energies $\mu \gg m_{W}$ the gauge boson  is effectively
massless, and, the gauge coupling decreases as we increase the scale, showing
QCD-like asymptotic freedom.  For energies $\mu \gtrsim m_{W}$, the mass of the
W becomes relevant and the coupling stops increasing.  The screening of the
gauge force due to the massive W boson is clear from the difference between the
perturbative and the non-perturbative result.  For $\mu\ll m_{W}$, the gauge
boson decouples and we effectively recover a scalar theory.

It is also interesting to explore the infrared structure of the running gauge
coupling.  For this reason, a coupling
$g_{\textrm{Y}}^{2}(\mu = 1/r) = r^{2}\dv{V_{\textrm{Y}}}{r}(r)$ from a
Yukawa-like potential \cite{langguth_monte_1986,fodor_simulating_1994},
$V_{\textrm{Y}}(r) \propto \frac{1}{r} e^{-mr}$ was fit to the lattice data in
\cref{fig:running_coupling}.  The results of the fit for the finest lattice is
shown in \cref{fig:running_coupling}, with the corresponding Debye
\textit{screening-mass} estimated\footnote{Since we are using a fixed form
$e^{-m/\mu}$ for the Yukawa potential the arbitrariness in the definition of the
renormalized running coupling, and the scale at which it is defined would lead
to different screening-masses.} as
$m_{\textrm{screen}}= 0.6243(28)m_{W}\approx 50~\si{GeV}$.


While the points the LPCP have the SM observables fixed to their physical
values, the BSM couplings remain as free parameters.   A particularly
interesting  objective would be to establish non-perturbative  bounds on the BSM
spectrum, which would require a scan over the whole parameter space.  Instead,
keeping the quartic couplings small, we explore the effect of changing the
`unbroken' hopping parameter $\kappa_{1}$ within the sector $(H_{2})$.

\begin{figure}[htb!]
\centering
  \includegraphics[width=0.49\textwidth]{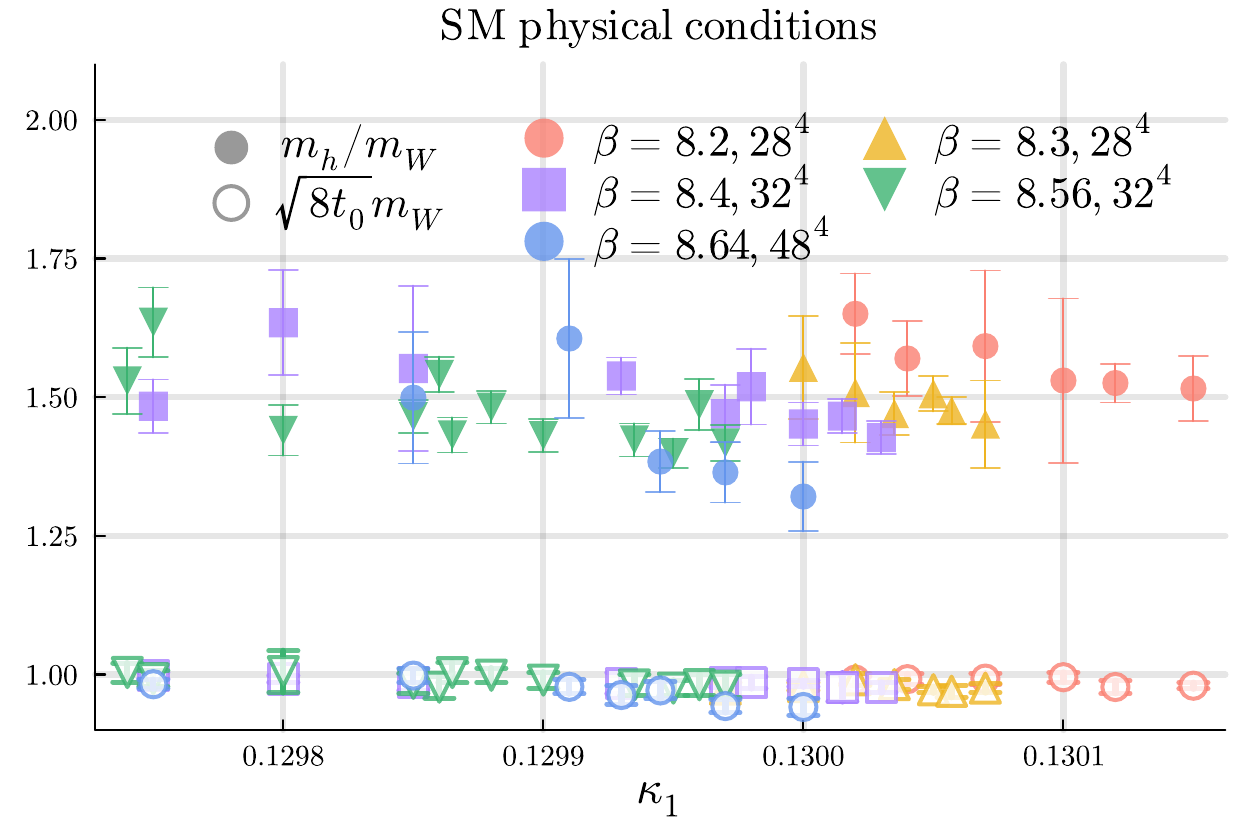}
  \includegraphics[width=0.49\textwidth]{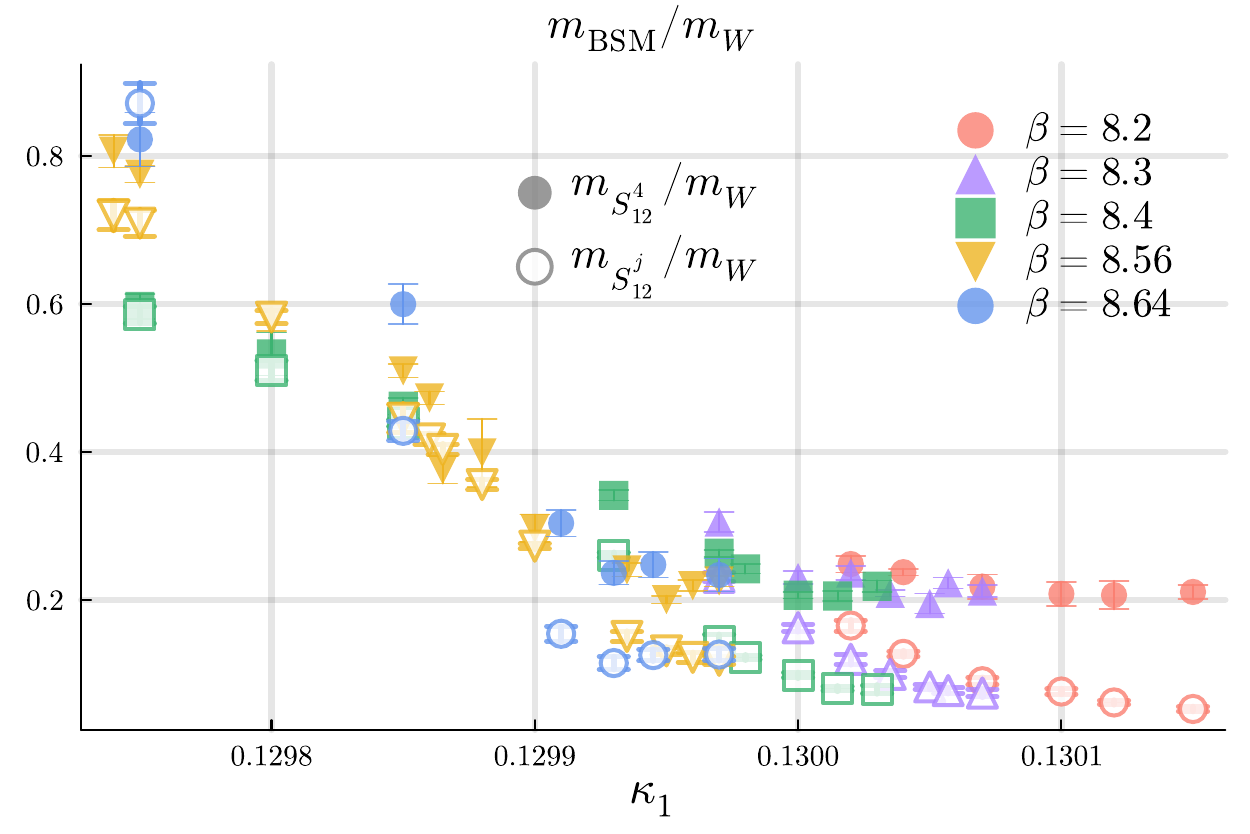}
  \caption{Left: Standard Model conditions for simulations at different
$\kappa_{1}$ with the remaining couplings defined in \cref{tab:LCP}. Right: mass
ratio between the BSM states,
$m_{S_{12}^{4}}=m_{H},~m_{S_{12}^{j}}=m_{A}=m_{H^{\pm}}$, and the W boson mass,
$m_{W}$, for the same points at different $\kappa_{1}$ below $(H_{12})$.}
  \label{fig:S_R_mBSMoW_BSM_scan}
\end{figure}

We performed simulations at different $\kappa_{1}$ values within sector
$(H_{2})$ for each $\beta$ in \cref{tab:LCP}.  The SM conditions were monitored
for each simulation, and the results are summarized in the left plot of
\cref{fig:S_R_mBSMoW_BSM_scan}.  Within the available precision the physical
conditions remain roughly unchanged by the change in $\kappa_{1}$.

The mass ratios of the BSM states with the W boson are shown in
\cref{fig:S_R_mBSMoW_BSM_scan} for all $\beta$ values in the LCPC.  The improved
precision allows us to observe the mass-gap between the $H$ and the $A,H^{\pm}$
states.  While the mass ratio $m_{H}/m_{W}$ develops a plateau when approaching
$\kappa_{1}^{c}$ from below, indicating a saturation to a finite value before
the transition into $(H_{12})$, the ratio  $m_{A}/m_{W}=m_{H^{\pm}}/m_{W}$ seems
to keep decreasing with increasing $\kappa_{1}$.

\begin{figure}[htb!]
  \centering
  \includegraphics[width=0.49\textwidth]{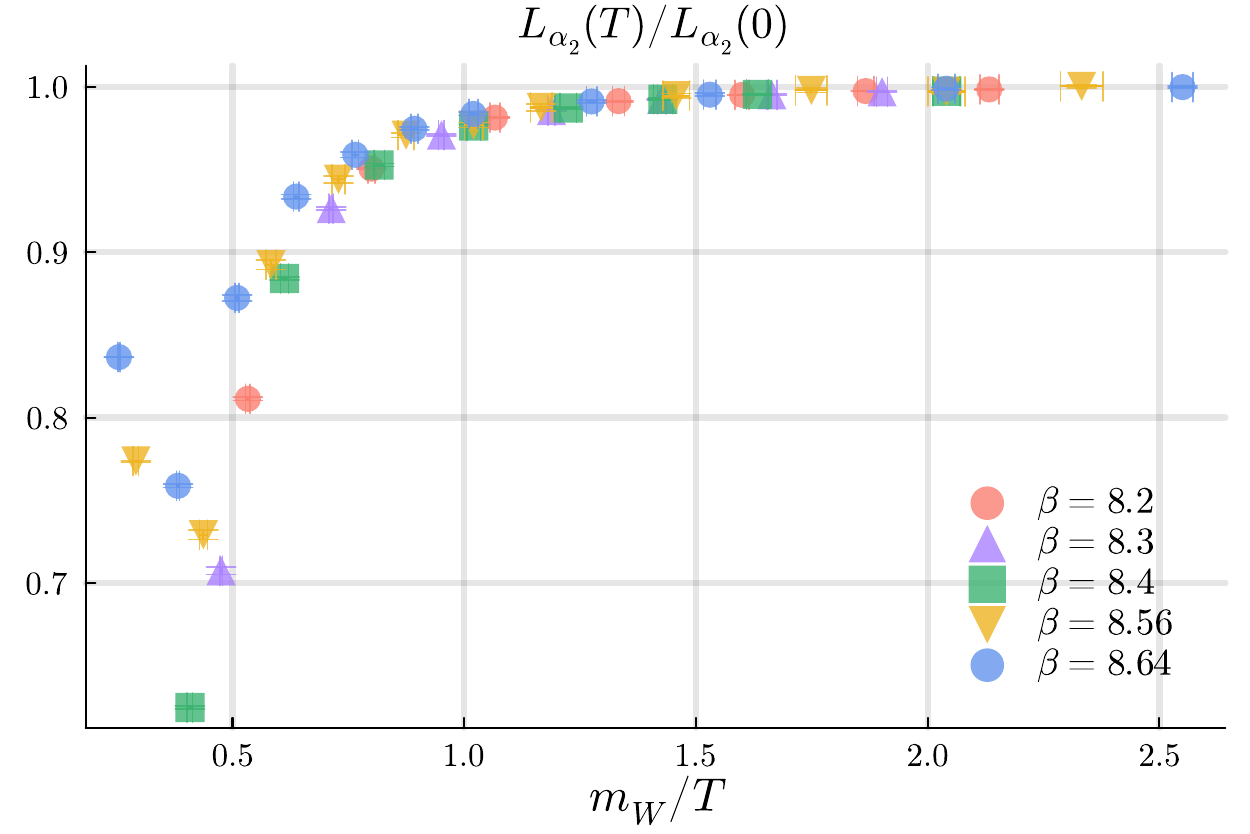}
  \includegraphics[width=0.49\textwidth]{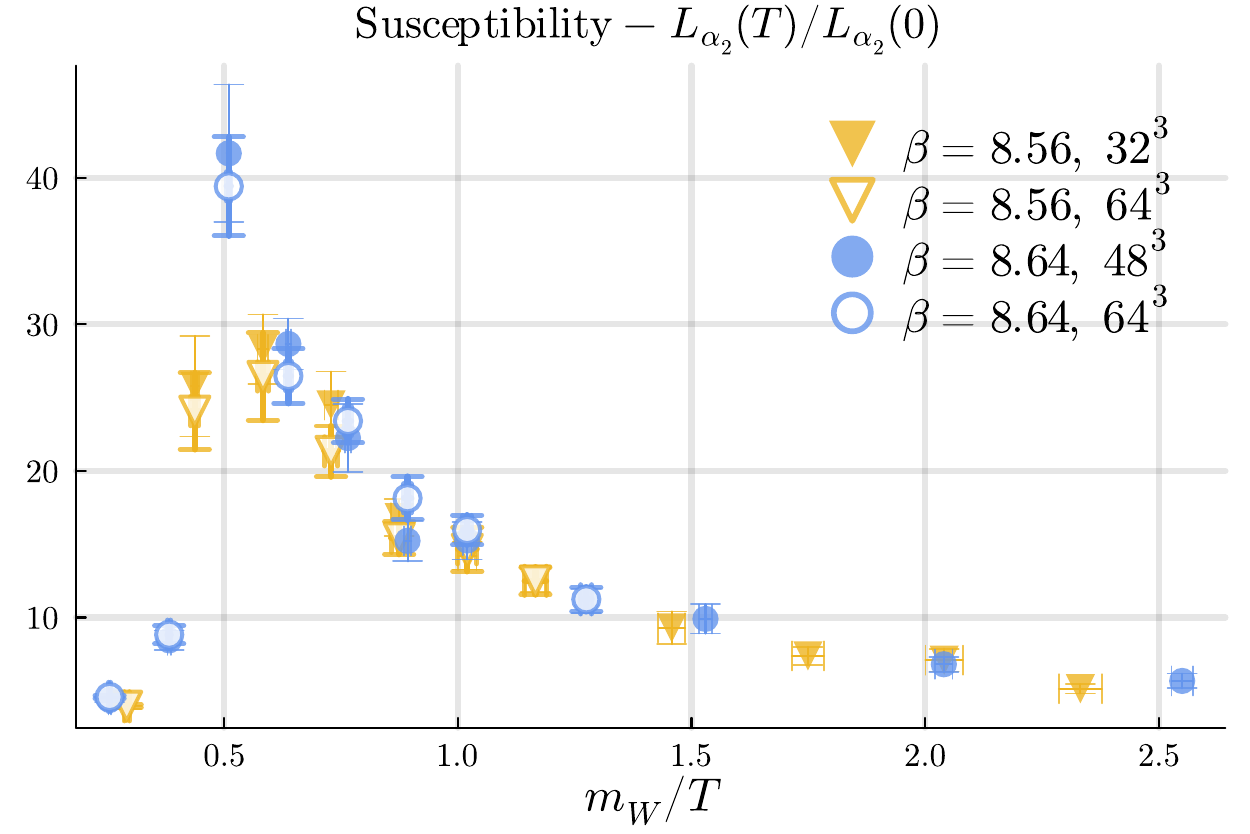}
\caption{Finite temperature dependence of the ratio
$\mathcal O(T,g)/\mathcal O(0,g)$ (left) and
$\chi_{\mathcal O}(T,g)/\chi_{\mathcal O}(0,g)$ (right) for
$\mathcal O = L_{\alpha_{2}}$. The susceptibility is shown only for the two
larges $\beta$ values and for different spatial volumes are shown.}
  \label{fig:ratio_vev_fintemp}
\end{figure}

In the following we discuss our work on the finite temperature transitions in
this model.  The points of the LPCP were simulated using asymmetric lattices,
with the temporal extent defining the physical temperature by $T=1/(aL_{4})$.
The `symmetry restoration' that deactivates the Higgs mechanism can be observed
in the global observable $L_{\alpha_{2}}$
\begin{equation}
  L_{\alpha_{2}} =\frac{1}{8V}\sum_{x,\mu}\Tr{\alpha_{2}^{\dagger}(x)U_{\mu}(x)\alpha_{2}(x+\hat\mu)},
\end{equation}
where $\alpha_{n}$ is the `angular' part of the quaternion Higgs field,
$\Phi_{n}=\rho_{n}\alpha_{n},~\rho_{n}\in\mathbb R,~\alpha_{n}\in SU(2)$
In the left plot of \cref{fig:ratio_vev_fintemp} the renormalized ratio
$L_{\alpha_{2}}(T,g)/L_{\alpha_{2}}(0,g)$ ($g$ denotes all couplings) is shown
as a function of the dimensionless ratio $ m_{W}/T$ with $am_{W}$ taken from the
zero temperature simulations.  Lattice artifacts can be seen at high energies
due to the use of very small temporal extents.

The departure of  $L_{\alpha_{2}}(T)/L_{\alpha_{2}}(0)$ from unity at high
temperatures signals the passage to the `symmetric' confinement region
$(H_{0})$, around $m_{W}/T_{c} \sim 0.5$, with $T_{c}$ the critical temperature.
In order to better resolve the transition point, and to understand the character
of this phase transition we compute the susceptibility by
$\chi(L) = L^{4}\left( \expval{L_{\alpha}^{2}} - \expval{L_{\alpha}}^{2} \right)$.
This is shown for $L_{\alpha_{2}}$ in the right plot of
\cref{fig:ratio_vev_fintemp} for the two finest lattices and different spatial
volumes.  The absence of volume dependence in the peak of the susceptibility
indicates that, similarly to the single Higgs case, the electroweak transition
in the weakly coupled 2HDM is a crossover.

\section{Conclusion} \label{sec:Conclusion}

We have studied the inert and custodial limit of the 2HDM along a line of
constant SM physics defined in the Higgs sector of the theory with the unbroken
$SU(2)$ custodial symmetry.  Both the Higgs-to-W mass ratio, and the
renormalized weak gauge coupling at the mass scale of the W boson were fixed to
their physical values in order to define a LPCP.

The running of the gauge coupling, computed with the gradient flow scheme along
the LPCP, defines a single curve for a large range of energies.  This was
compared with the one-loop massless scheme at high energies, and with a Yukawa
potential at low energies.

A scan in the BSM sector was performed with constant SM physics, allowing to
probe different regions of the parameter space.  We have found that very light
($m_{H}\sim 0.2m_{W}$) BSM scalar states are realizable within the LPCP tuned at
weak quartic couplings.

The finite temperature transition was also studied along the LPCP.  The results
show a transition occurring around $m_{W}/T_{c} \sim 0.5$ for the finest $\beta$
values. However, only the finest two lattices allow the observation of a well
defined peak in the susceptibility.  The lack of volume dependence in the latter
indicates a smooth crossover behavior at small quartic couplings for the
electroweak phase transition, which is in agreement with most of the
perturbative tree-level predictions.

The use of small BSM quartic couplings have been particularly helpful in this
work. As indicated before, they make the BSM sector roughly independent of the
SM physics, and prevent the need of retuning the LPCP when scaning the BSM
sector.  While this is an advantage from the computational perspective, there is
no fundamental reason for the choice of small couplings, and a complete study is
necessary in the future.  In fact, $\order{1}$ quartic couplings are thought to
be required for a strong first-order electroweak phase transition.  Tree-level
results \cite{basler_strong_2017,bernon_new_2018} indicate the need of  large
mass splitting within the BSM scalar sector for the occurrence of a strong
first-order transition.  This condition directly relates to an enhanced value of
the inter-flavour Higgs quartic couplings, namely $\eta_{3}$.

\section*{Acknowledgements}

The authors acknowledge financial support from the Generalitat Valenciana (genT program CIDEGENT/2019/040), Ministerio de Ciencia e Innovacion (PID2020-113644GB-I00), the Academic Summit Project -- NSTC 112-2639-M-002-006-ASP of Taiwan.
GC, CJDL, and AR acknowledge the CSIC-NSTC exchange grant 112-2927-I-A49-508.
CJDL acknowledges the financial support from NSTC project 112-2112-M-A49-021-MY3.
Similarly, GWSH, CJDL and MS acknowledge the support from the NSTC project 112-2639-M-002-006-ASP.

\bibliography{references}

\end{document}